\documentclass[aps,preprint,superscriptaddress,nofootinbib,preprintnumbers]{revtex4-1}

\usepackage{amsmath,amssymb,amsfonts}
\usepackage{bm, slashed, soul}
\usepackage{color,graphicx}
\usepackage{hyperref}

\usepackage{xcolor}
\definecolor{red}{rgb}{0.9, 0,0}

\newcommand{\hzgver}{h Z\gamma}
\newcommand{\hzg}{h \to Z\gamma}
\newcommand{\hgll}{h \to \gamma Z \to \gamma \ell^+\ell^-}

\newcommand{\glgll}{gg \to \gamma Z \to \gamma \ell^+\ell^-}
\newcommand{\glhgll}{gg \to h \to \gamma Z \to \gamma \ell^+\ell^-}
\newcommand{\ctilde}{\tilde{c}}
\newcommand{\thg}{\theta_\gamma}
\newcommand{\thz}{\theta_Z}
\newcommand{\phiz}{\phi_Z}
\newcommand{\mz}{m_Z}
\newcommand{\Gz}{\Gamma_Z}

\newcommand{\sigbr}[4]{ \langle #2^#1| #3 | #4^#1 \rangle}

\newcommand{\beq}{\begin{equation}}
\newcommand{\eeq}{\end{equation}}
\newcommand{\bea}{\begin{align}}
\newcommand{\eea}{\end{align}}

\allowdisplaybreaks[1]
\begin{document}

\title{Probing CP Violation in $\hzg$ with Background Interference}

\def\affcornell{Department of Physics, LEPP, Cornell University, Ithaca, NY 14853 \vspace*{8pt}}

\author{Marco Farina}
\affiliation{\affcornell}
\author{Yuval Grossman}
\affiliation{\affcornell}
\author{Dean J. Robinson}
\affiliation{Department of Physics, University of California, Berkeley, CA 94720, USA}
\affiliation{Ernest Orlando Lawrence Berkeley National Laboratory,
University of California, Berkeley, CA 94720, USA}

\begin{abstract}
We show that the parity of the $h Z\gamma$ vertex can be probed by interference
between the gluon fusion Higgs production, $gg \to h \to \gamma Z \to \gamma \ell^+\ell^-$,
and the background, $gg \to \gamma Z \to \gamma \ell^+\ell^-$, amplitudes.
In the presence of a parity violating $hZ\gamma$ vertex, this interference alters
the kinematic distribution of the leptons and photon compared to Standard Model (SM)
expectations. For a Higgs with SM-sized width and couplings, we
find that the size of the effect enters at most at the $10^{-2}$ level.
Such a small effect cannot be seen at the LHC, even with futuristic
high luminosities. Should there exist other broader scalar particles with
larger production cross-section times branching ratio to $Z\gamma$, then the parity
structure of their $Z\gamma$ couplings can be probed with this technique.
\end{abstract}

\maketitle

\section{Introduction}
Measurement of the Higgs boson couplings is a primary search channel for New Physics (NP). The magnitude of the Higgs couplings can be tested against Standard Model (SM) predictions via measurement of Higgs production cross sections, branching fractions, and overall decay rate. On the other hand, these couplings may also encode NP in exotic parity structure, which cannot be probed by Higgs branching fractions, but instead requires a more sophisticated analysis. For example, angular distributions in the four body final state processes $h \to ZZ^* \to  \ell^+\ell^-\ell^+\ell^-$ or $h \to WW^* \to  \ell^+\nu \ell^-\bar\nu$ probe the parity structure of the $hZZ$ and $hWW$ couplings~\cite{Aad:2013xqa,Chatrchyan:2013mxa,Chatrchyan:2013iaa}.

In this work we concentrate on the $\hzgver$ vertex. To lowest order this vertex has two effective couplings: $c$, which is parity even, and $\ctilde$, which is parity odd (see definitions in eq.~\eqref{eqn:HEF} below).
In the SM, $c$ is generated at one-loop level, while $\ctilde = 0$ to a very good approximation because it violates parity. Since the leading order SM contribution arises only at one-loop, this raises the possibility that one-loop NP effects can be comparable to the SM terms, producing P-violating effects in the $\hzg$ decay at the $\mathcal{O}(1)$ level. By contrast, $\mathcal{O}(1)$ parity violating effects are already ruled out in the $h \to ZZ$ and $h \to WW$ channels, that are dominated by tree level SM contributions.

Several prior studies have suggested several different approaches to  probe of the structure of the $\hzgver$ vertex. In Ref.~\cite{Korchin:2013ifa} the parity structure is probed through a forward-backward asymmetry in the $\hgll$ decay, exploiting the parity violating coupling of the $Z$ to leptons. This asymmetry requires the presence of a non-negligible relative strong phase in the $\ctilde$ coupling. This strong phase is generated by on-shell $b$-quark loops, and thus the asymmetry is typically suppressed by $m_b/m_t$.  Another approach is to exploit the interference between one-loop $h \to Z\gamma^* \to \ell^+\ell^-\ell^+\ell^-$ and the much larger, tree-level $h \to ZZ^*\to \ell^+\ell^-\ell^+\ell^-$ amplitude~\cite{Chen:2014gka,Chen:2015iha}. The challenge here is to distinguish P-violating effects arising from P-odd $hZ\gamma$ operator and a P-odd $hZZ$ operator, which is not excluded from arising at the one-loop level. Similarly, interference of $hZ\gamma$ and $h\gamma\gamma$ P-even and P-odd operators in $h \to \ell^+\ell^-\gamma$ may generate a forward-backward asymmetry~\cite{Chen:2014ona}. Since the $Z$ is dominantly on-shell but the photon is virtual,  this interference is suppressed by $\Gz/\mz$. Moreover, one cannot distinguish contributions of the P-odd $\hzg$ and $h \to \gamma\gamma$ operators. Finally, one might use converted photons, in a similar fashion to what was proposed for $h \to \gamma\gamma$~\cite{Kroll:1955ip,Cottingham:1995sh,Voloshin:2012tv,Bishara:2013vya}. The drawback in this approach is the currently limited experimental ability to resolve the leptonic angular structure of the conversion.

In this work we explore another option to probe the parity structure of the $\hzgver$ vertex: Exploitation of the interference between  $gg \to \hgll$ and its background $\glgll$. In particular, we construct, in a general model-independent fashion, an angular kinematic observable whose oscillatory probability distribution is amplitude-modulated and phase-shifted by interference effects in the presence of parity violation. SM Higgs-background interference effects have been previously considered for the $h \to \gamma \gamma$ channel \cite{Dixon:2003yb, Dixon:2008xc,Martin:2013ula}, though not in the context of searches for exotic parity structure.

While in principle our method provides a new handle to unambiguously probe the $\hzgver$ vertex, one may expect that performing the analysis in practice will be challenging. In the first instance, the gluon fusion process, $\glgll$, is characteristically suppressed by $(\alpha_s/4\pi)^2~\mbox{pdf}(gg)/\mbox{pdf}(q\bar{q}) \sim 0.02$ compared to the dominant $q\bar{q} \to \gamma Z \to \gamma \ell^+\ell^-$ background. The $gg$ and the $q\bar{q}$ channels albeit add incoherently, so that the background interference effect is not spoiled in principle. Nevertheless, the observables discussed in this paper will therefore be buried in this $q\bar{q}$ background. Moreover, the narrow Higgs width, $\Gamma_h \simeq 4$~MeV in the SM, significantly reduces the phase space over which interference effects with the background can be large.

Using numerical simulations we estimate that the background interference effect is present at the $10^{-2}$ to $10^{-3}$ level for the amplitude modulation and phase shift respectively. Unfortunately, these effects are small enough that they cannot be seen at the LHC, even with maximal parity violation in the $\hzgver$ vertex and a futuristic luminosity of $3$~ab$^{-1}$. If, however, there is a new scalar with either a larger gluon fusion production cross-section times branching ratio to $Z\gamma$ or a larger total width, for instance, a singlet scalar or a heavy Higgs arising from a two-Higgs doublet model, then the parity structure of this coupling may be probed by this method.

\section{Framework}
\subsection{Higgs Effective Theory}
Keeping terms up to dimension $d=5$, the effective theory of interest for the gluon-fusion Higgs production channel $gg \to \hgll$ is
\begin{equation}
	\label{eqn:HEF}
	\mathcal{L}_{\rm h} = \frac{c}{v}~h\,F_{\mu\nu} Z^{\mu\nu} + \frac{\ctilde}{2v}~h\,F_{\mu\nu}\tilde{Z}^{\mu\nu} + \frac{c_g}{v}~h\,G^a_{\mu\nu}G^{a\mu\nu}~.
\end{equation}
Here $F$, $Z$, and $G^a$ denote respectively the photon, $Z$ and gluon field strengths, $v=246$ GeV is the electroweak vacuum expectation value, and the dual field strength is defined as usual as $\tilde{X}^{\mu\nu} \equiv \epsilon^{\mu\nu\alpha\beta}X_{\alpha\beta}$. We also assume the Higgs is a $J^{PC} = 0^{++}$ state. The leading order SM expressions for the couplings are explicitly
\begin{align}
{c_g}_{\rm SM}
	&= \frac{\alpha_s}{4 \pi} \sum_q \mathcal{A}_{1/2}^H \bigg[\frac{4 m_q^2}{m_h^2},\frac{4 m_q^2}{m_Z^2}\bigg] \notag\\
	& \simeq \frac{\alpha_s}{4 \pi} ( -0.344 - 0.005 i) \,\, , \notag \\
c_{\rm SM}
	&= \frac{\alpha}{4 \pi s_W} \bigg(\mathcal{A}_{1}^H\bigg[\frac{4 m_W^2}{m_h^2},\frac{4 m_W^2}{m_Z^2}\bigg]+ \sum_f N_c \frac{Q_f v_f}{c_W}\mathcal{A}_{1/2}^H \bigg[\frac{4 m_f^2}{m_h^2},\frac{4 m_f^2}{m_Z^2}\bigg] \bigg) \notag\\
	&\simeq \frac{\alpha}{4 \pi s_W} (5.508 - 0.004 i)~,\notag\\
\ctilde_{\rm SM}
	& \simeq 0~. \label{eqn:CSM}
\end{align}
Expressions for the functions $\mathcal{A}^H_{1,1/2}$ can be found in Ref.~\cite{Gunion:1988st} (see also \cite{Bergstrom:1985ih}).

Provided the digluon invariant mass $s < 2m_{W}$, rescattering phases in the $\hzgver$ vertex arise only from on-shell $b$ quark loops, and are therefore suppressed by $m_b/m_t$. This is the origin of  the small imaginary parts in eqs~\eqref{eqn:CSM}.  Consequently, for $s \sim m^2_h$ -- the invariant mass interval of interest in this paper -- we may neglect these small strong phases, and assume that $c$, $\ctilde$, and $c_g$ are real.

The parity odd and parity even contributions add incoherently in the $\hzg$ rate, so that $\Gamma_{\hzg} \propto c^2 + \ctilde^2$. We define
\begin{equation}
	\label{eqn:MUZGD}
	\mu_{Z\gamma} \equiv \frac{\Gamma_{\hzg}}{\Gamma^{\rm SM}_{\hzg}} = \frac{c^2 + \ctilde^2}{c_{\rm SM}^2}\,.
\end{equation}
We assume that there are no NP effects in the Higgs-gluon couplings, that is, $c_g = {c_g}_{\rm{SM}}$ and the only source of P violation is the  $hF\tilde{Z}$ operator. It is then convenient to define
\begin{equation}
	\label{eqn:CHSD}
	\xi \equiv \tan^{-1}(\ctilde/c)~,
\end{equation}
and rewrite the $\hzgver$ couplings as 
\begin{equation}
c = \mu_{Z\gamma} c_{\rm SM}\cos\xi\,, \qquad \ctilde = \mu_{Z\gamma}c_{\rm SM}\sin\xi\,.
\end{equation}
In the case that the partial width $\Gamma_{\hzg}$ is SM-like, i.e. $\mu_{Z\gamma} = 1$, then $\xi$ alone encodes the parity structure: All NP effects manifest in $\xi$, with $\xi = 0$ in the SM to a very good approximation.

\subsection{Amplitude Factorization}
Since we are interested in processes with only two external gluons, the color structure in any amplitude must be proportional to the identity. We therefore drop color indices henceforth, treating the gluons formally as photons, and consider only color-stripped amplitudes: appropriate traces and color factors are kept implicit.

Both the Higgs channel, $\glhgll$, and background, $\glgll$ process, factorize into a $2\to2$ scattering, a $Z$ propagator, and a $1\to2$ $Z$ decay. Close to the $Z$ mass shell, the propagator is well-approximated by a Breit-Wigner form. With respect to the effective theory~\eqref{eqn:HEF}, the Higgs channel has the diagrammatic form
\begin{align}
	[\mathcal{M}_{\rm h}]^{\lambda_1\lambda_2}_{\lambda\tau_-\tau_+}
	&  = \qquad \parbox[c]{9cm}{
	\includegraphics[scale=1]{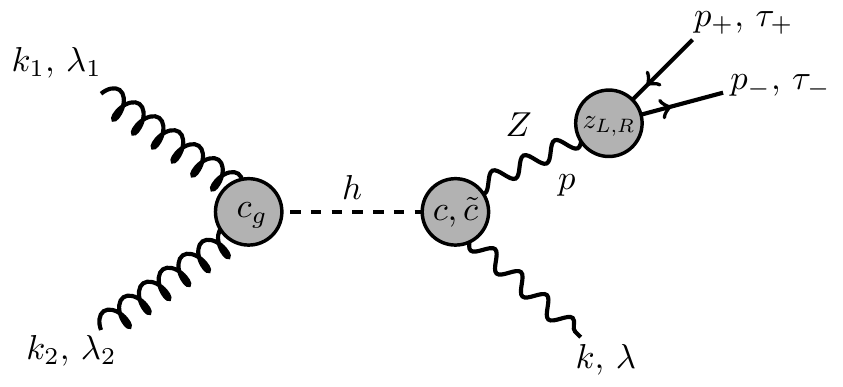}
	}\notag\\
	& \equiv  \frac{g_{\mu\nu} - p_\mu p_\nu / \mz^2}{p^2 - \mz^2 + i \mz \Gz} [\mathcal{M}_{2 \to 2, \rm{h}}]^{\lambda_1 \lambda_2\mu}_{ \lambda} [\mathcal{M}_{1 \to 2}]^{\nu}_{\tau_-\tau_+}~.\label{eqn:FFHC}
\intertext{Similarly, the background amplitude is given by}
	[\mathcal{M}_{\rm bg}]^{\lambda_1\lambda_2}_{ \lambda\tau_-\tau_+}
	& = \frac{g_{\mu\nu} - p_\mu p_\nu / \mz^2}{p^2 - \mz^2 + i \mz \Gz} [\mathcal{M}_{2 \to 2, \rm{bg}}]^{\lambda_1 \lambda_2\mu}_{ \lambda} [\mathcal{M}_{1 \to 2}]^{\nu}_{\tau_-\tau_+}~,\label{eqn:FFBGC}
\end{align}
Here $\mathcal{M}_{1\to2}$ is the $Z \to \ell^+\ell^-$ amplitude, while $\mathcal{M}_{2 \to 2, \rm{h}}$ ($\mathcal{M}_{2 \to 2, \rm{bg}}$) are the $2\to2$ Higgs channel (background) scattering amplitudes. The momenta and helicities of the gluons (photon) are respectively denoted by $k_i$ and $\lambda_i$ ($k$ and $\lambda$), $p$ denotes the $Z$ momentum, while $p_\pm$ are the lepton momenta and $\tau_\pm = \pm$ are their respective spins. Hereafter we neglect the lepton masses, so spins correspond to definite helicity states. The $Z$ chiral couplings to leptons are denoted by $z_{L,R}$.

In the massless lepton limit, the $Z \to \ell^+\ell^-$ amplitude, $[\mathcal{M}_{1 \to 2}]^{\nu}_{\tau_-\tau_+}$, annihilates the $p_\mu p_\nu / \mz^2$ piece of the $Z$ propagator via the lepton equations of motion. We may then make use of the $Z$ polarization completeness relation for any $p^2 \not=0$
\begin{equation}
	 -g_{\mu\nu} + p_\mu p_\nu/p^2 = \sum_\kappa \epsilon^{\kappa*}_\mu (p)\epsilon^{\kappa}_\nu(p)~,
\end{equation}
to rewrite eqs.~\eqref{eqn:FFHC} and \eqref{eqn:FFBGC} into the factorized form
\begin{equation}
	\label{eqn:AF}
	[\mathcal{M}_{\rm{h},\rm{bg}}]^{\lambda_1\lambda_2}_{ \lambda \tau_-\tau_+} =  \frac{1}{p^2 - \mz^2 + i\mz\Gz}\sum_{\kappa = -,0,+}[{\mathcal{M}_{2 \to 2}}_{\rm{h},\rm{bg}}]^{\lambda_1 \lambda_2}_{ \lambda \kappa} [\mathcal{M}_{1 \to 2}]^\kappa_{\tau_-\tau_+}~,
\end{equation}
where $\kappa$ is the helicity of the $Z$, defined in a consistently chosen frame (for instance, the digluon center-of-mass frame). Hence, the amplitude factorizes into two Lorentz invariant factors and a Breit-Wigner propagator.

In the $\Gz \to 0$ limit, the Breit-Wigner denominator in the squared amplitude
\begin{equation}
	\label{eqn:NZL}
	\frac{1}{\big|p^2 - \mz^2 + i m_Z \Gz\big|^2} \to \pi\frac{1}{m_Z\Gz}\delta( p^2 - \mz^2)~,
\end{equation}	
which ensures the $Z$ is on-shell. That is, the full $2 \to 3$ amplitude is well-approximated with an internal on-shell $Z$, up to corrections of order $\Gamma^2_Z/\mz^2 \ll 1$. Hereafter we shall therefore replace the squared Breit-Wigner propagator by the $\Gz \to 0$ limit $\delta$-function, and enforce an on-shell $Z$. It should be emphasized, however, that the helicity amplitude analysis that follows below relies only on the factorized form \eqref{eqn:AF}, which holds typically up to $m_\ell/\mz$ corrections, and independently from the $\Gz \to 0$ limit. Off-shell $Z$ effects merely alter the overall normalization of the square amplitude and the phase space volume, but do not affect the helicity amplitude structure encoded by eq. \eqref{eqn:AF}.

\subsection{Phase space coordinates}
In general, a $2\to 3$ phase space is described by five variables. One of them is the dilepton invariant mass $(p_+ + p_-)^2 \simeq \mz^2$ in the narrow $Z$ limit. Correspondingly to eq.~\eqref{eqn:AF}, the remaining four-dimensional phase space may be partitioned into two Lorentz invariant phase spaces: the phase spaces of the $2\to2$ and the $1\to2$ processes.

\begin{figure}[ht]
	\begin{tabular*}{0.7\textwidth}{@{\extracolsep{\fill}}cc}
	\includegraphics[scale=1]{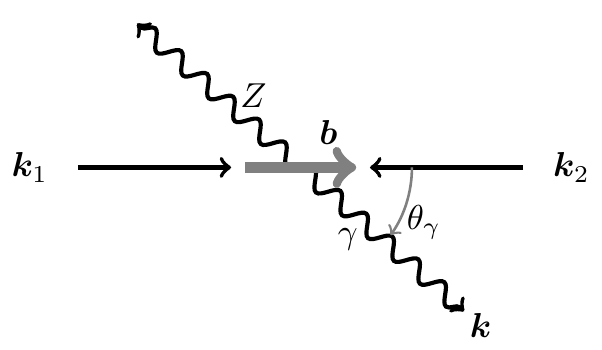} & \includegraphics[scale = 1]{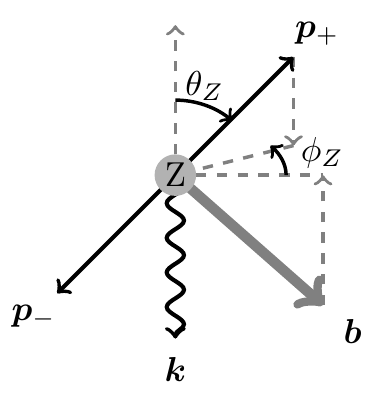}
	\end{tabular*}
	\caption{Definition of polar angles. Left: photon polar angle, $\thg$, with respect to beam line in gluon center of mass frame. Right: Positron polar angles, $\thz$ and $\phiz$, with respect to photon and beam line, $\bm{b}$, in $Z$ rest frame.}
	\label{fig:PAD}
\end{figure}

We assume only longitudinal gluon boosts with respect to the beam line. The $2\to2$ phase space is then described conveniently by the digluon invariant mass $(k_1 + k_2)^2 = s$, and the photon polar angle, $\thg$, defined with respect to the beam axis, $\bm{b}$, in the gluon center of mass frame (see Fig.~\ref{fig:PAD}). The remaining two variables encode the $1\to 2$ phase space, and are conveniently chosen to be the polar and azimuthal angle of the positively charged lepton in the $Z$ rest frame, $\thz$ and $\phiz$, defined with respect to the photon momentum and beam axis in that frame (see Fig.~\ref{fig:PAD}). In appendix~\ref{app:KV} we write down the construction of the $\{s,\thg\}$ and $\{\thz,\phiz\}$ coordinates in terms of $2\to2$ Mandelstam and other Lorentz invariants.

Applying eqs.~\eqref{eqn:AF} and \eqref{eqn:NZL}, it follows from these choices that the full differential cross-section
\begin{equation}
	\label{eqn:DCS}
	\frac{d\sigma(s, \thg;\thz,\phiz) }{d(\cos \thg) d(\cos\thz) d\phiz}= \frac{(s - \mz^2)}{2^{11}\pi^3s^2}\frac{\big|\mathcal{M}(s, \thg;\thz,\phiz)\big|^2}{\mz\Gz}~,
\end{equation}
where
\begin{equation}
	\label{eqn:FFSA}
	\big|\mathcal{M}(s, \thg;\thz,\phiz)\big|^2 \equiv \sum_{\lambda_i,\lambda,\tau_\pm}\bigg|\sum_{\kappa = -,0,+}[\mathcal{M}_{2 \to 2}]^{\lambda_1 \lambda_2}_{ \lambda \kappa}(s,\thg) [\mathcal{M}_{1 \to 2}]^\kappa_{\tau_-\tau_+}(\thz,\phiz)\bigg|^2~.
\end{equation}

\section{Parity Observable}
\subsection{\texorpdfstring{$1\to2$}{12} Amplitudes}
We now construct explicit expressions for the $1\to2$ helicity amplitudes of eq.~\eqref{eqn:FFSA} and examine their properties under discrete transformations. Amplitudes are computed with spinor-helicity methods. Our particular choices for the polarizations and reference momenta are shown in Appendix~\ref{app:SPC}. With these choices, the $Z\to \ell^+\ell^-$ amplitudes can be shown to take the simple form
\begin{align}
	[\mathcal{M}_{1\to 2}]^+_{+-}(\thz,\phiz) & = \sqrt{2}\mz z_L e^{i\phiz} \sin^2(\thz/2)~,\notag\\
	[\mathcal{M}_{1\to 2}]^+_{-+}(\thz,\phiz) & = -\sqrt{2}\mz z_R e^{i\phiz} \cos^2(\thz/2)~,\notag\\
	[\mathcal{M}_{1\to 2}]^-_{+-}(\thz,\phiz) & = -\sqrt{2}\mz z_Le^{-i\phiz} \cos^2(\thz/2)~,  \notag\\
	 [\mathcal{M}_{1\to 2}]^-_{-+}(\thz,\phiz) & = \sqrt{2}\mz z_Re^{-i\phiz} \sin^2(\thz/2)~,  \notag\\
	[\mathcal{M}_{1\to 2}]^0_{+-}(\thz,\phiz) & = -\mz z_L \sin\thz~,\notag\\
	 [\mathcal{M}_{1\to 2}]^0_{-+}(\thz,\phiz) & = -\mz z_R \sin\thz~,\label{eqn:12Z}
\end{align}
and all other amplitudes are zero. Here $z_{L,R}$ are the $Z$ chiral couplings to leptons. Note that these amplitudes are $j=1$ Wigner d-matrix functions. In the massless lepton limit, the non-zero $1\to 2$ amplitudes require $\tau_- = -\tau_+$, so hereafter we shall always write
\begin{equation}
	[\mathcal{M}_{1\to 2}]^{\kappa}_{\tau} \equiv [\mathcal{M}_{1\to 2}]^{\kappa}_{\tau,-\tau}~.
\end{equation}

Let us now examine the discrete P, C and CP transformations of $[\mathcal{M}_{1\to 2}]^{\kappa}_{\tau}$. The index $\kappa$ is defined by the choices~\eqref{eqn:PCV} to be the $Z$ helicity in the digluon center-of-mass frame. $\kappa$ changes sign under parity, P, as does the lepton helicity $\tau$. This is equivalent to $z_L \leftrightarrow z_R$ and $\phiz \to -\phiz$, the latter arising because the sense of the azimuthal twist of the lepton momenta around their parent changes sign under parity. In other words, $\phiz$ acts like a weak phase under parity conjugation. From eqs.~\eqref{eqn:12Z}, one may explicitly check that
 \begin{equation}
	\label{eqn:ZP}
	[\mathrm{P} \mathcal{M}_{1\to 2}]^{\kappa}_{\tau}(\thz,\phiz) =  [\mathcal{M}_{1\to 2}]^{-\kappa}_{-\tau}(\thz,\phiz) = [\mathcal{M}_{1\to 2}]^\kappa_{\tau}(\thz,-\phiz)\Big|_{z_L\leftrightarrow z_R}~.
\end{equation}
Charge conjugation C switches $\tau_+$ and $\tau_-$, and hence sends $\tau \to - \tau$. Equivalently ${z_L\leftrightarrow z_R}$, $\thz \to \pi - \thz$, and $\phiz \to \pi + \phiz$. That is,
\begin{equation}
	\label{eqn:ZC}
	[\mathrm{C} \mathcal{M}_{1\to 2}]^{\kappa}_{\tau}(\thz,\phiz) = [\mathcal{M}_{1\to 2}]^{\kappa}_{-\tau}(\thz,\phiz) = [\mathcal{M}_{1\to 2}]^{\kappa}_{\tau}(\pi-\thz,\pi +\phiz)\Big|_{z_L\leftrightarrow z_R}~,
\end{equation}
which can similarly be explicitly checked from eqs.~\eqref{eqn:12Z}. Finally, combining eqs. \eqref{eqn:ZP} and \eqref{eqn:ZC}, under CP
\begin{align}
	[\mathrm{CP} \mathcal{M}_{1\to 2}]^{\kappa}_{\tau}(\thz,\phiz) & = [\mathcal{M}_{1\to 2}]^{-\kappa}_{\tau}(\thz,\phiz) = [\mathcal{M}_{1\to 2}]^{\kappa}_{\tau}(\pi - \thz,\pi-\phiz) \notag\\
	& = (-1)^\kappa[\mathcal{M}_{1\to 2}]^{\kappa}_{\tau}(\pi - \thz,-\phiz)~,	\label{eqn:ZCP}
\end{align}
where the second line here follows from the special explicit form of the $1\to2$ amplitudes~\eqref{eqn:12Z}.

\subsection{\texorpdfstring{$2\to2$}{22} Amplitudes}

With reference to the effective theories~\eqref{eqn:HEF} and to eqs.~\eqref{eqn:FFHC}, the Higgs channel $2 \to 2$ amplitude factorizes into $gg \to h$ and $\hzg$ amplitudes, connected by a propagator. One finds that
\begin{equation}
	\label{eqn:22H}
	[\mathcal{M}_{2\to2,\rm{h}}]^{\lambda_1 = \lambda_2}_{\lambda = \kappa}(s,\thg) = \frac{s(s - \mz^2)}{v^2}\frac{i \mu_{Z\gamma}\,{c_g}_{\rm SM}\,c_{\rm SM} }{s - m_h^2 + im_h\Gamma_h} e^{i \kappa\xi}~,
\end{equation}
and all other amplitudes are zero. Note that the amplitude does not depend on $\thg$ since the Higgs is a scalar. This amplitude carries a phase arising from the Higgs propagator, which acts as a strong phase under parity conjugation. On the other hand, $\xi$ is a weak phase. That is, the parity relation for the non-zero amplitudes is
\begin{equation}
	\label{eqn:HCPR}
	[\mathrm{P} \mathcal{M}_{2\to2,\rm{h}}]^{\lambda_1\lambda_2}_{\lambda\kappa}(s) = [\mathcal{M}_{2\to2,\rm{h}}]^{-\lambda_1-\lambda_2}_{-\lambda-\kappa}(s) = [\mathcal{M}_{2\to2,\rm{h}}]^{\lambda_1\lambda_2}_{\lambda \kappa}(s)\Big|_{\xi \leftrightarrow -\xi}~.
\end{equation}

Computation of the background $2 \to 2$ helicity amplitudes requires evaluation of box diagrams with light internal quarks. Note that these amplitudes may contain non-negligible strong phases, arising from on-shell internal degrees of freedom, relative to $c$, $\ctilde$ and $c_g$.  Explicit results for the helicity amplitudes as functions of Mandelstam variables are available in Ref.~\cite{Ametller:1985ro}. However, the $Z$ polarization basis chosen therein is not necessarily commensurate with the choices~\eqref{eqn:PCV} that lead to the especially simple results in eqs.~\eqref{eqn:12Z}. Instead, for the purposes of this general analysis, we leave the background amplitudes in the abstract form $[\mathcal{M}_{2 \to 2,\rm{bg}}]^{\lambda_1 \lambda_2}_{ \lambda \kappa}(s,\thg)$, and note only that all phases contained in these amplitudes are strong phases under parity conjugation. In Appendix~\ref{app:PHA}, by taking the heavy quark limit, we show that their parity transformation is
\begin{equation}
	\label{eqn:BGP}
	[\mathrm{P} \mathcal{M}_{2\to2,\rm{bg}}]^{\lambda_1\lambda_2}_{\lambda\kappa}(s,\thg) = [\mathcal{M}_{2 \to 2,\rm{bg}}]^{-\lambda_1 -\lambda_2}_{ -\lambda -\kappa}(s,\thg) = -(-1)^\kappa [\mathcal{M}_{2 \to 2,\rm{bg}}]^{\lambda_1 \lambda_2}_{ \lambda \kappa}(s,\thg)~.
\end{equation}
Comparing to \eqref{eqn:HCPR}, the parity transformation for the Higgs channel amplitudes also has parity $-(-1)^\kappa$, since those amplitudes are trivially zero in the case $\kappa = 0$. Eq. \eqref{eqn:BGP} implies that the $\kappa=0$ background channel is odd under parity, while the $\kappa = \pm 1$ $Z$ channels are even. Note finally that the C transformation on all $\mathcal{M}_{2 \to 2}$ amplitudes is trivial.

\subsection{Discrete Symmetries of the Differential Cross-section}
Combining the CP transformations~\eqref{eqn:ZCP}, \eqref{eqn:HCPR} and \eqref{eqn:BGP}, the full amplitude has a CP transformation
\begin{equation}
	\label{eqn:FACPR}
	[\mathrm{CP} \mathcal{M}]^{\lambda_1\lambda_2}_{\lambda\tau}(s,\thg;\thz,\phiz) = [\mathcal{M}]^{-\lambda_1-\lambda_2}_{-\lambda\tau}(s,\thg;\thz,\phiz) = -[\mathcal{M}]^{\lambda_1\lambda_2}_{\lambda\tau}(s,\thg;\pi-\thz,-\phiz)\bigg|_{\xi\to-\xi}~.
\end{equation}
The corresponding polarized differential cross-sections~\eqref{eqn:DCS}, after marginalizing over $\thz$ phase space, therefore obey
\begin{equation}
	\frac{d [\sigma]^{-\lambda_1-\lambda_2}_{-\lambda\tau}(s,\thg;\phiz)}{d\cos\thg d\phiz} = \frac{d [\sigma]^{\lambda_1\lambda_2}_{\lambda\tau}(s,\thg;-\phiz)}{d\cos\thg d\phiz}\bigg|_{\xi\to-\xi}~,
\end{equation}
because $\int_{-1}^1 f(\pi - \thz) d\cos\thz = \int_{-1}^1 f(\thz) d\cos\thz$ for any integrable $f$. It follows that the unpolarized differential cross-section
\begin{equation}
	\label{eqn:DCSCP}
	\frac{d \sigma(s,\thg;\phiz)}{d\cos\thg d\phiz}  = \frac{d \sigma(s,\thg;-\phiz)}{d\cos\thg d\phiz} \bigg|_{\xi \to -\xi}\,.
\end{equation}
I.e. it is invariant under the simultaneous weak phase transformations $\xi \to - \xi$ and $\phiz \to -\phiz$. Similar application of the C transformation
\begin{equation}
	\label{eqn:FACR}
	[\mathrm{C} \mathcal{M}]^{\lambda_1\lambda_2}_{\lambda\tau}(s,\thg;\thz,\phiz) = [\mathcal{M}]^{\lambda_1\lambda_2}_{\lambda-\tau}(s,\thg;\thz,\phiz) = [\mathcal{M}]^{\lambda_1\lambda_2}_{\lambda\tau}(s,\thg;\pi-\thz,\pi+\phiz)\bigg|_{z_L\leftrightarrow z_R}~,
\end{equation}
yields
\begin{equation}
	\label{eqn:DCSC}
	\frac{d \sigma(s,\thg;\phiz)}{d\cos\thg d\phiz} = \frac{d \sigma(s,\thg;\pi+\phiz)}{d\cos\thg d\phiz} \bigg|_{z_L\leftrightarrow z_R}\,.
\end{equation}
I.e. the unpolarized differential cross-section is invariant under the simultaneous transformation $\phiz \to \pi + \phiz$ and $z_L \leftrightarrow z_R$.

\subsection{CP sensitive observable}
We now proceed to show that the parity violating parameter $\xi$ manifests as a phase in the probability distribution of the kinematic observable $\phiz$. First, eqs.~\eqref{eqn:12Z} and \eqref{eqn:22H} together imply that the $Z$ helicity, $\kappa$, encodes the weak phase structure of the amplitude. That is, we can rewrite the amplitude into the form
\begin{equation}
	\label{eqn:AWPS}
	\mathcal{M}^{\lambda_1 \lambda_2}_{ \lambda \tau}(\thz,\phiz) =  z_\tau \bigg[\sum_{\kappa=\pm}[\mathcal{A}^{\rm h}_\kappa]^{\lambda_1\lambda_2}_{\lambda \tau}(\thz)e^{ i\kappa(\xi + \phiz)} + 	\sum_{\kappa=0,\pm}[\mathcal{A}^{\rm bg}_\kappa]^{\lambda_1\lambda_2}_{\lambda \tau}(\thz)e^{ i\kappa\phiz }\bigg] ~.
\end{equation}
where the $s,\thg$ arguments have been omitted for brevity, and we have explicitly extracted the chiral coupling $z_{\tau= \pm} = z_{L,R}$ as well as weak phase $\phiz$, $\xi$ dependence from the $1\to2$ amplitudes~\eqref{eqn:12Z}. We may further rewrite~\eqref{eqn:AWPS} more compactly as
\begin{equation}
	\label{eqn:AWPSA}
	\mathcal{M}^{\lambda_1 \lambda_2}_{ \lambda \tau}(\thz,\phiz) =  z_\tau \sum_{k = h_\pm,0,\pm}[\mathcal{A}_k]^{\lambda_1\lambda_2}_{\lambda \tau}(\thz)\exp\big\{ i\eta_k\big\}~,
\end{equation}
where the $\mathcal{A}_{h_\pm}$ terms arise from the Higgs channel~\eqref{eqn:22H} -- note that $[\mathcal{A}_{h_\pm}]^{\lambda_1\lambda_2}_{\lambda\tau} = 0$ for $\lambda = \mp$ respectively --  and the $\mathcal{A}_{0,\pm}$ terms are the background amplitude contributions. The $\eta_{k}$ are the corresponding weak phases: $\eta_{h_\pm,0,\pm} = \pm(\xi + \phiz)$, $0$, and $\pm\phiz$ respectively. Comparing eq.~\eqref{eqn:AWPSA} with the parity relations~\eqref{eqn:ZP}, \eqref{eqn:HCPR} and \eqref{eqn:BGP}, one deduces that the $\mathcal{A}_{k}$ transform under parity as
\begin{equation}\
	\label{eqn:AKPR}
	[{\mathrm{P}}(\mathcal{A}_k)]^{\lambda_1\lambda_2}_{\lambda \tau} = -(-1)^k[\mathcal{A}_k]^{\lambda_1\lambda_2}_{\lambda \tau}\,,
\end{equation}
with the notational understanding that $(-1)^{k = \pm,h_\pm} = -1$ and $(-1)^{k = 0} = 1$.

The CP relation~\eqref{eqn:DCSCP} implies that the differential cross-section $d\sigma/d\cos\thg d\phiz$ is the average of itself and its weak phase conjugation,
\begin{equation}
	d\sigma/d\cos\thg d\phiz = \big[d\sigma/d\cos\thg d\phiz + d\sigma/d\cos\thg d\phiz\big|_{\eta_k \to -\eta_k}\big]/2\,. 
\end{equation}
Applying this result to the explicit form \eqref{eqn:AWPS}, one may then show that
\begin{equation}
	\frac{d \sigma(s,\thg;\phiz)}{d\cos\thg d\phiz} = \frac{(s - \mz^2)}{s^2 \mz\Gz} \Bigg\{\sum_{k}\mathcal{B}_{kk}(s,\thg)+  \! \sum_{k\not=l}\mathcal{B}_{kl}(s,\thg) \cos\big[\eta_k - \eta_l \big]\Bigg\}~,
\end{equation}
wherein
\begin{equation}
	\mathcal{B}_{kl}(s,\thg) = \frac{2^{-\delta_{kl}}}{2^{11}\pi^3}\sum_{\lambda_{i},\lambda,\tau}z_\tau^2\int_{-1}^1 d\cos\thz \mbox{Re}\Big\{ [\mathcal{A}_k \mathcal{A}^*_l]^{\lambda_1\lambda_2}_{\lambda \tau}(s,\thg;\thz) \Big\}~.
\end{equation}
One then further deduces from the C relation~\eqref{eqn:DCSC} that the unpolarized marginal differential cross-section, integrated over an interval $s \in I$ containing the Higgs peak $s = m_h^2$, must have the form
\begin{align}
	 \frac{d \sigma^I}{d\phiz}
	 = \frac{1}{\mz\Gz}\Bigg\{ (z_L^2 + z_R^2)\bigg[&\sum_{k=h_\pm,0,\pm} \mathcal{C}^I_{kk} + [\mathcal{C}^I_{h_++} + \mathcal{C}^I_{h_--}]\cos(\xi) \notag\\
	& + [\mathcal{C}^I_{+-} + \mathcal{C}^I_{-+}]\cos(2\phiz)  +  [\mathcal{C}^I_{h_+-} + \mathcal{C}^I_{h_-+}]\cos (\xi + 2\phiz)\bigg]\notag\\
	  + (z_L^2 - z_R^2)\bigg[& [\mathcal{C}^I_{0-} + \mathcal{C}^I_{0+}]\cos(\phiz)  +  [\mathcal{C}^I_{h_+0} + \mathcal{C}^I_{h_-0}]\cos (\xi + \phiz)\bigg]\Bigg\}\,,
\end{align}
in which $d \sigma^I/d\phiz \equiv \int_I (d\sigma/d\phiz) ds$, and
\begin{equation}
	\mathcal{C}^I_{kl} = 2^{-\delta_{kl}}\int_I ds \frac{s - \mz^2}{2^{11}\pi^3s^2} \sum_{\lambda_{i},\lambda}\int_{-1}^1\int_{-1}^1 d\cos\thz d\cos\thg \mbox{Re}\Big\{ [\mathcal{A}_k \mathcal{A}^*_l]^{\lambda_1\lambda_2}_{\lambda+}(s,\thg;\thz) \Big\}~.
\end{equation}
Note that the $\xi$ dependent terms arise manifestly from Higgs and background interference terms. The terms proportional to $z_L^2 - z_R^2$ arise from the combination of the parity violating $Z$-lepton chiral couplings and interference between the $\kappa=0$ and $\kappa = \pm1$ background channels. The parity relations \eqref{eqn:AKPR} for the $\mathcal{A}_{0,\pm}$ imply that these terms are parity odd, so that they vanish under the $d\cos\thg$ integral. This is to be expected, as the background QCD process is not parity violating. Hence, in summary 
\begin{multline}
	\frac{d \sigma^I}{d\phiz}  = \frac{1}{\mz\Gz}\Bigg\{ (z_L^2 + z_R^2)\bigg[\sum_{k=h_\pm,0,\pm} \mathcal{C}^I_{kk} + [\mathcal{C}^I_{h_++} + \mathcal{C}^I_{h_--}]\cos(\xi) \\
	 + [\mathcal{C}^I_{+-} + \mathcal{C}^I_{-+}]\cos(2\phiz)  +  [\mathcal{C}^I_{h_+-} + \mathcal{C}^I_{h_-+}]\cos (\xi + 2\phiz)\bigg]\,.\label{eqn:MRP}
\end{multline}
Eq.~\eqref{eqn:MRP} is the main analytical result of this paper. In the presence of background interference the parity structure of $\hzg$ can be probed through the phase structure of $d\sigma^I/d\phiz$. In particular, the differential cross-section is asymmetric about $\phiz =0$ if and only if $\xi \not= 0$. If $\mathcal{C}^I_{kl}$ are known, then fitting this function permits in principle extraction or bounding of $\xi$.

The $\cos(\xi + 2\phiz)$ term in eq.~\eqref{eqn:MRP} further implies that a quadrant-type asymmetry can be generated by integrating over the $\phiz$ quadrants
\begin{equation}
	 \rm{I} = [0,\pi/2]~, \quad \rm{II} = [\pi/2, \pi]~, \quad \rm{III} = [\pi,3\pi/2]~, \quad \mbox{and} \quad \rm{IV} = [3\pi/2, 2\pi]~,
\end{equation}
and taking their alternating sum. In particular,
\begin{equation}
	\label{eqn:SPZ}
	\Sigma_{\phiz} \equiv \frac{1}{\sigma}\int_{-\rm{I} + \rm{II} - \rm{III} + \rm{IV}}\bigg(\frac{d \sigma^I}{d\phiz} \bigg) d\phiz = \frac{2}{\pi} \frac{[\mathcal{C}^I_{h_+-} + \mathcal{C}^I_{h_-+}]\sin(\xi) }{\sum_{k} \mathcal{C}^I_{kk} + [\mathcal{C}^I_{h_++} + \mathcal{C}^I_{h_--}]\cos(\xi) }\,,
\end{equation}
is non-zero if and only if $\xi \not=0$.

\subsection{Simulations}
It remains to estimate the $\mathcal{C}^I_{kl}$ coefficients of eq. \eqref{eqn:MRP}. This is achieved with a privately modified version of the $\glgll$ process found in \texttt{MCFM-6.8} (process \#300) \cite{Campbell:2011bn}, customized to include an interfering set of $\glhgll$ helicity amplitudes. The relative strong phase between the extant \texttt{MCFM} background amplitudes and the Higgs channel couplings \eqref{eqn:CSM} is hard to extract from Refs. \cite{Gunion:1988st} and \cite{Campbell:2011bn}. We have checked, however, that any potential mismatch of conventions, leading to an extra strong phase, is of small numerical significance for the purposes of estimating the $\mathcal{C}^I_{kl}$ coefficients: It introduces at most an extra signed $\mathcal{O}(1)$ factor. We choose the Higgs peak region $I =\{ \sqrt{s} \in (124,128)~\mbox{GeV}\}$. All simulations are generated for a $pp$ collider running at $14$~TeV, with photon transverse cuts $p_T > 20$~GeV and $|\eta| < 2.5$, and dilepton invariant mass $m_{\ell\ell} \in (66,116)$~GeV.

For the sake of brevity, let us rewrite the integrated version of eq.~\eqref{eqn:MRP} in a compact form
\begin{equation}
	\label{eqn:MRPC}
	\frac{d \sigma^I}{d \phiz}(\phiz;\xi) = a_0 + a_2\cos(2\phiz) + b_0 \cos(\xi) + b_2\cos(2\phiz + \xi)~.
\end{equation}	
This may be further rewritten into an SM-normalized form
\begin{equation}
	\label{eqn:MRPC2}	
	\frac{d \sigma^I}{d \phiz} = \frac{\sigma^I_{\rm SM}}{2\pi}\frac{1}{1 + b_0/a_0}\bigg[ 1+ a_2/a_0 \cos(2\phiz) + b_0/a_0 \cos(\xi) +  b_2/a_0 \cos(2\phiz + \xi)\bigg]\,,
\end{equation}
where $\sigma^I_{\rm SM}$ is the SM ($\xi =0$) $\glgll$ cross-section on the interval $I$, including Higgs interference effects. We may now in principle extract the relative coefficients $a_2/a_0$, $b_0/a_0$ and $b_2/a_0$ by generating $\phi_Z$ distributions over $I$ for various values of $\xi$ and $\phi_Z \in \{-\pi,\pi\}$, and fitting $a_{0,2}$ and $b_{0,2}$ to eq.~\eqref{eqn:MRPC}. However, very high statistics are required to sufficiently sample over the narrow SM Higgs width, $\Gamma_h \simeq 4$~MeV, in order to extract the $b_{0,2}$ coefficients with satisfactory precision.

Instead, let us apply the Higgs coupling and width rescalings
\begin{equation}
	\label{eqn:HWR}
	c \to \zeta c~,\qquad \ctilde \to \zeta \ctilde~,\qquad \mbox{and} \qquad \Gamma_h \to \zeta^2 \Gamma_h~.
\end{equation}
The pure Higgs $\mathcal{C}_{h_\pm h_\pm}$ terms, which exclusively arise in $a_0$, are dominated by on-shell Higgs contributions $\sim (c^2 + \ctilde^2)/\Gamma_h$, and therefore are invariant under this rescaling. The pure background $\mathcal{C}_{\pm\pm}$ and $\mathcal{C}_{00}$ terms are invariant by definition. Hence $a_2$ and $a_0$ are unchanged by the transformation \eqref{eqn:HWR}. However, the Higgs-background interference terms $\mathcal{C}_{h_{\pm} \pm}$ -- the $b_{0,2}$ coefficients -- are enhanced by a $\zeta$ factor. The larger Higgs width and enhanced interference effects together admit faster numerical convergence of the coefficients of eq.~\eqref{eqn:MRPC}: The $b_{0,2}$ coefficients extracted for $\zeta \gg 1$ may then be rescaled by $\zeta^{-1}$ to determine their SM values.

While we shall use the $\zeta$ rescaling \eqref{eqn:HWR} as a numerical tool, it should be noted that $\zeta$ itself can be measured or bounded: Constraints on interference effects far off the Higgs mass shell may also be used to bound the total Higgs width, and hence $\zeta$ (see e.g.~\cite{Caola:2013yja,Campbell:2013wga,Campbell:2013una,Campbell:2014gha,Kauer:2015pma}). Currently, bounds from the $h \to 4\ell$ channel imply $\zeta \lesssim 3$ \cite{Altas:2014do,Khachatryan:2014iha}. The $\hzg$ partial width itself is invariant under the rescaling \eqref{eqn:HWR}, but the current upper bound on $\sigma\times\mbox{Br}[ \glhgll]$ is approximately an order of magnitude above the SM value~\cite{Chatrchyan:2013vaa,Aad:2014fia}. There is, therefore, still some room for NP enhancements of the $\hzg$ rate itself. Collectively, possible NP effects in the Higgs total and partial $Z\gamma$ width can be encapsulated by the rescalings \eqref{eqn:MUZGD} and \eqref{eqn:HWR}, viz.
\begin{equation}
	\label{eqn:HPWR}
	c \to \mu_{Z\gamma}^{1/2}\,\zeta\, c~,\qquad \ctilde \to \mu_{Z\gamma}^{1/2}\,\zeta\, \ctilde~,\qquad \mbox{and} \qquad \Gamma_h \to \zeta^2\,\Gamma_h~.
\end{equation}
with $\zeta\lesssim 3$ and $\mu_{Z\gamma} \lesssim 10$.

\begin{figure}[t]
	\includegraphics[width=8cm]{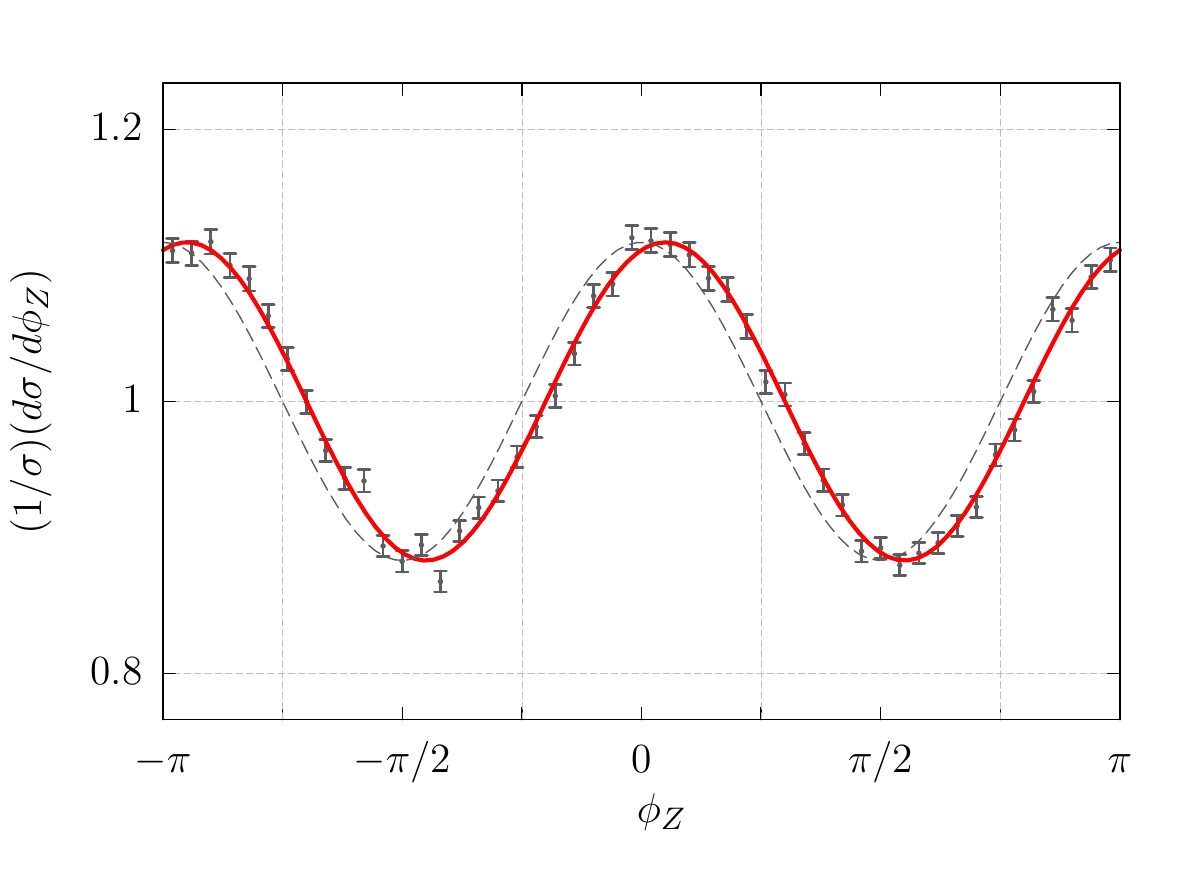}
	\caption{The $\phiz$ distribution and best-fit curve (red) in the gluon fusion channel for $\sqrt{s} \in (124,128)$~GeV, for $\xi = \pi/2$ and rescaling factor $\zeta^2 = 10^{3}$. The Higgs couplings have been increased by a further factor of three in order to enhance the visibility of the shift due to $\xi \not=0$. This shift can be seen in the displacement of the best fit curve with respect to the cosine $\sim \cos(2\phiz)$ (dashed line).}
	\label{fig:PZP}
\end{figure}

Under this rescaling approach, we generate $\phi_Z$ distributions for $\zeta = 10$, and $\xi = 0$, $\pi/4$, $\pi/3$, $\pi/2$ and $\pi$. After rescaling the $b_{0,2}$ coefficients to their SM values, one finds in this manner
\begin{align}
	a_2/a_0 & \equiv \frac{\mathcal{C}^I_{+-} + \mathcal{C}^I_{-+}}{ \sum_{k} \mathcal{C}^I_{kk}}  = 0.143 \pm 0.001 \notag\\
	b_0/a_0 &  \equiv \frac{\mathcal{C}^I_{h_++} + \mathcal{C}^I_{h_--}}{ \sum_{k} \mathcal{C}^I_{kk}}  = (6.61 \pm 0.08)\times 10^{-3} \notag\\
	b_2/a_0 & \equiv \frac{\mathcal{C}^I_{h_+-} + \mathcal{C}^I_{h_-+}}{\sum_{k} \mathcal{C}^I_{kk}} = -(0.92 \pm 0.08)\times 10^{-3}~. \label{eqn:DCR}
\end{align}
with cross-section $\sigma^I_{\rm SM} \simeq 2.33$ fb. The errors here are purely statistical in origin. The $\cos(2\phiz + \xi)$ and $\cos(\xi)$ coefficients are non-zero at high statistical confidence.  A typical \texttt{MCFM} $\phi_Z$ distribution, for $\zeta^2 = 10^3$ and $\mu_{Z\gamma} = 3$, together with its best fit curve are shown in Fig.~\ref{fig:PZP}, in which the expected shifted cosine can be seen.

\begin{figure}[t]
	\includegraphics[width=8cm]{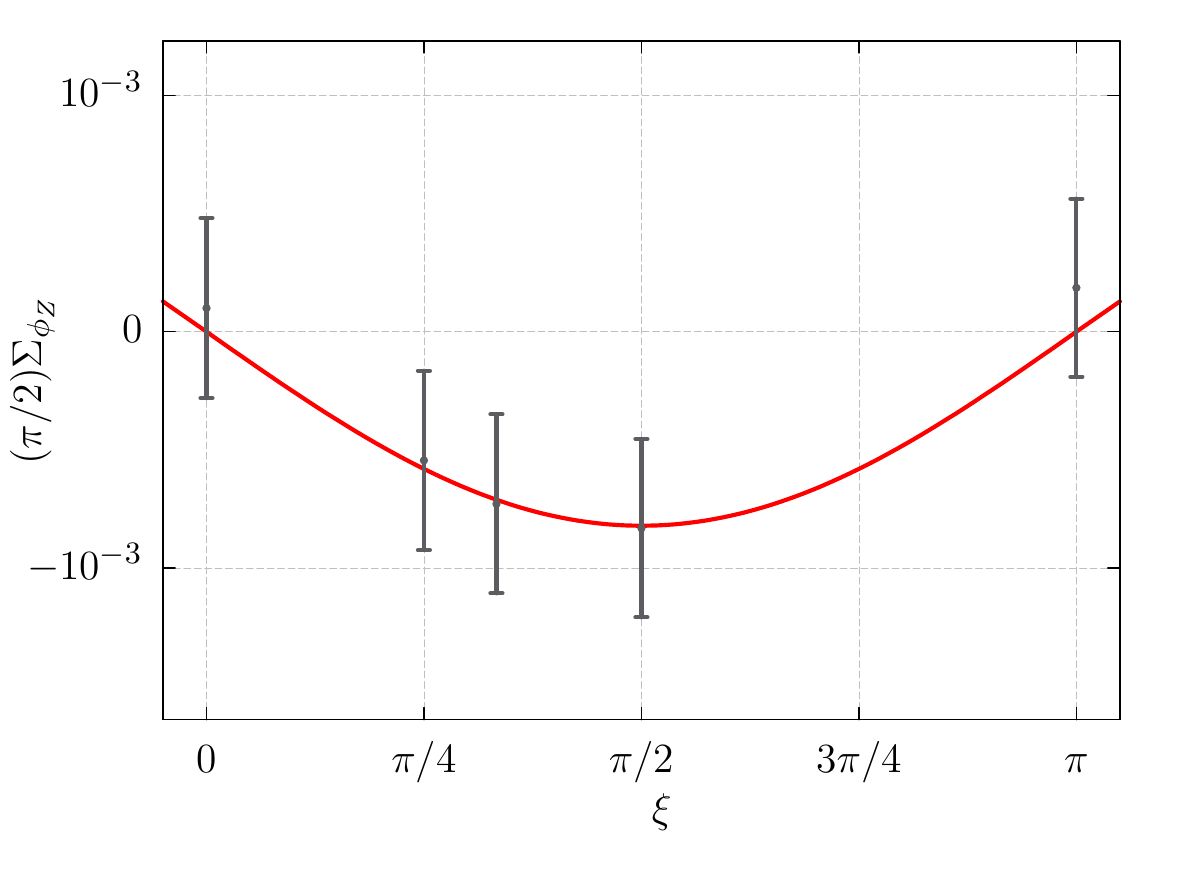}
	 \caption{The integrated asymmetry $\Sigma_{\phiz}$ for $\sqrt{s} \in (124,128)$~GeV and SM Higgs width and coupling magnitudes ($\zeta = 1$). The best fit curve of form $\sim\sin(\xi)$ is also shown (red).}
	\label{fig:IA}
\end{figure}

The asymmetry, $\Sigma_{\phiz}$, for the same set of $\xi$ values is also computed, and shown in Fig.~\ref{fig:IA}. Fitting to the expected dependence $(2/\pi)b_2\sin(\xi)/(a_0 + b_0\cos(\xi) )$ in eq.~\eqref{eqn:SPZ} with the assumption $b_0/a_0 \ll 1$, one finds that the best fit
\begin{equation}
	b_2/a_0 \equiv \frac{\mathcal{C}^I_{h_+-} + \mathcal{C}^I_{h_-+}}{\sum_{k} \mathcal{C}^I_{kk}} = -(0.84 \pm 0.07)\times 10^{-3}~,
\end{equation}
which is consistent with the fit from the $\phi_Z$ distributions.

\begin{figure}[t]
	\includegraphics[width=8cm]{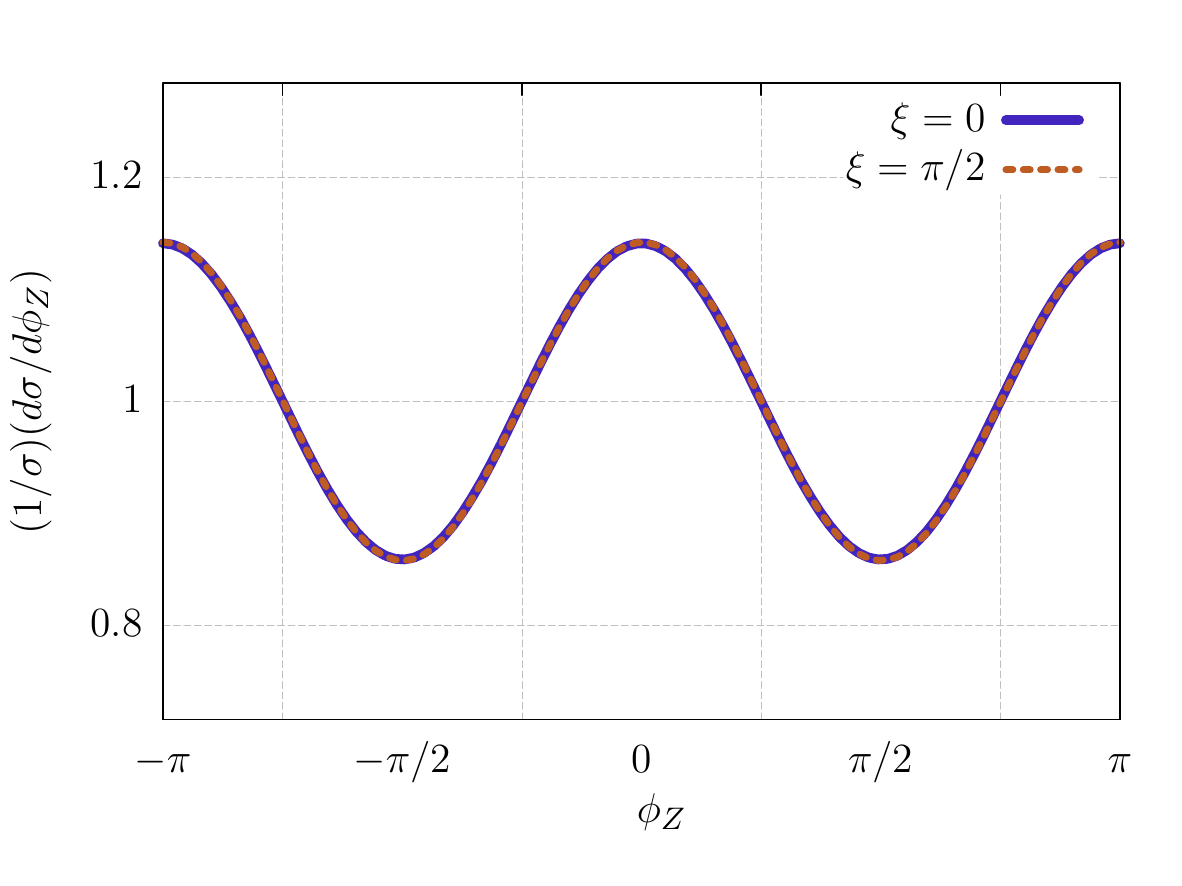}
	\includegraphics[width=8cm]{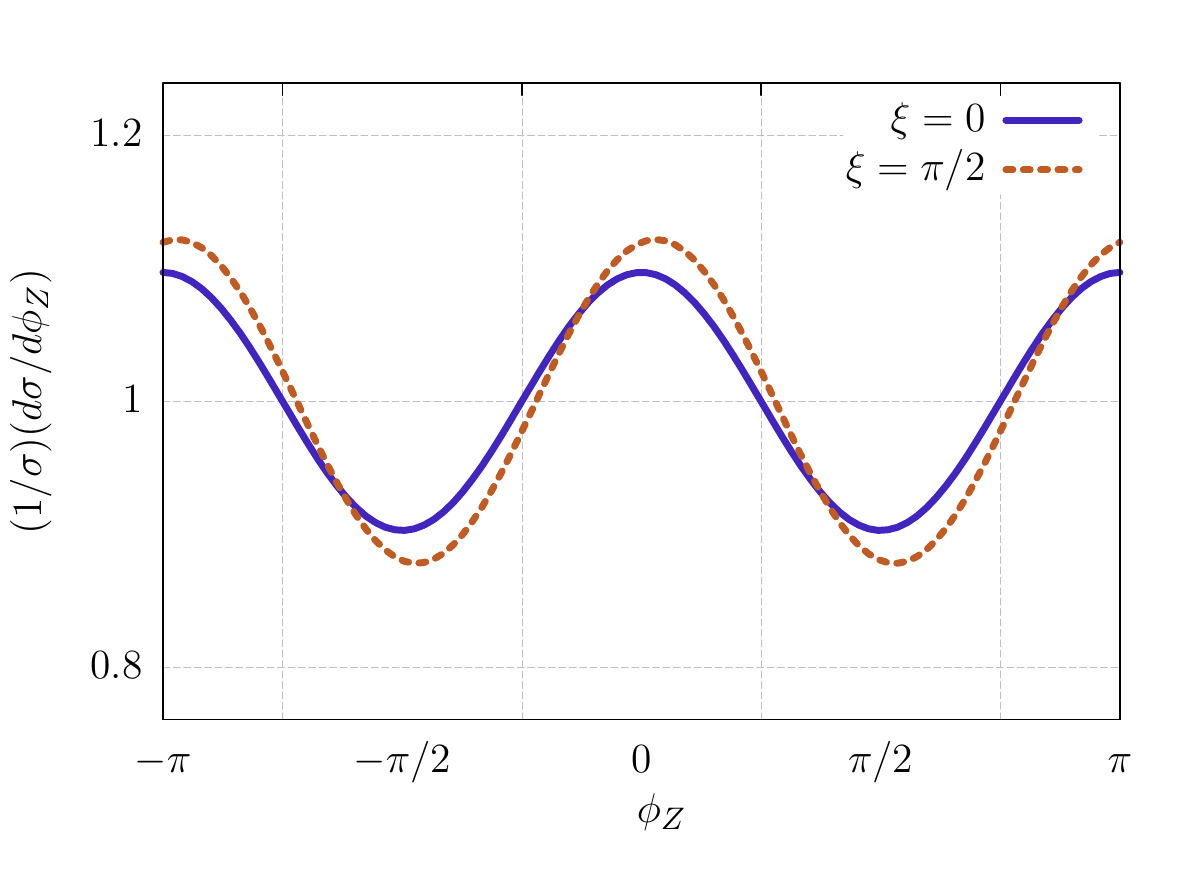}
	\caption{Left: The $\phiz$ best-fit SM distribution in the gluon fusion channel for $\sqrt{s} \in [124,128]$~GeV, for $\xi = \pi/2$ (gold) and $\xi = 0$ (blue) normalized to their respective cross-sections. Right: The same distribution, but for a width and coupling rescaling $\zeta = 30$. Compared to the SM result, the $\xi=\pi/2$ curve is shifted to the right, and has a larger oscillation amplitude.}
	\label{fig:PZF}
\end{figure}

We see from the results~\eqref{eqn:DCR} that the $\cos(\xi)$ coefficient in eq.~\eqref{eqn:MRPC2} dominates the $\cos(2\phiz + \xi)$ coefficient by an order of magnitude, $\sim 1\%$ and $\sim 0.1\%$ respectively in the SM. In Fig.~\ref{fig:PZF} we show the best-fit $d \sigma/d \phiz$ function for the zero and maximal parity violating cases $\xi = 0$ and $\xi = \pi/2$, with $\zeta = 1$. For comparison we also show $d\sigma/d\phiz$ for a $\zeta = 30$ scenario, which could correspond to a hypothetical Higgs-like particle with larger width and couplings. We see there that the $\cos(\xi)$ term manifests for $\xi \not = 0$ as a modulation of the oscillation amplitude of $d\sigma/d\phiz$ compared to SM expectations. It also rescales the overall cross-section from the expected SM value. Extracting or bounding $\xi$ by searching for these $\cos\xi$ term effects therefore requires computation of these SM expectations to sub-percent level, at which higher order QCD corrections likely become important. Hence this approach is susceptible to large theory errors.

In contrast, for $\xi \not = 0$ we see in Fig.~\ref{fig:PZF} that the $\cos(2\phiz + \xi)$ term manifests as a phase shift in $d\sigma/d\phiz$ with respect to the SM cosine. Equivalently $\Sigma_{\phiz} \not =0$. The SM expectation that such a phase shift is zero, or $\Sigma_{\phiz} =0$, holds to high loop order. Hence, even though searching for phase-shift or quadrant asymmetries is more difficult experimentally than searching for $\cos(\xi)$ term effects -- $b_2/a_0 \ll b_0/a_0$ and hence more statistics are required -- it is theoretically much cleaner.

Including the incoherent $q\bar{q}$ background, neglected so far in this discussion, this phase shift effect is expected to be further suppressed to the $\mathcal{O}(10^{-5})$ level, requiring exquisite measurement of the $pp \to Z\gamma$ differential cross-section.  Since $\sigma^{gg}/\sigma^{q\bar{q}} \sim 0.02$, then from eqs.~\eqref{eqn:DCR} we expect $\sigma_{\rm SM} \sim 100$~fb on the interval $I$. A simple estimate of the required luminosity to detect $\Sigma_{\phiz} \not=0$ can be obtained by noting that measurement of $\Sigma_{\phiz}$ is a counting experiment on the four $\phiz$ quadrants. For a large number of total events, $N$, the statistical error in $\Sigma_{\phiz}$ is at leading order $1/\sqrt{N}$.  In order to reject the SM hypothesis ($\xi=0$) at $95\%$ confidence, the required precision is then $2/\sqrt{N} \simeq 10^{-5}$, implying a required integrated luminosity $\gtrsim 10^8$~fb$^{-1}$. Note further that we have neglected possible systematic errors, which renders this estimate to be an optimistic one. Consequently, even in the proposed high luminosity future of LHC runs, with luminosity $\sim 3$~ab$^{-1}$, there will be insufficient statistics to achieve sensitivity to $\mathcal{O}(10^{-5})$ effects, and hence there is no plausible sensitivity to $\xi \not= 0$ for a Higgs with SM-sized couplings.

\section{Conclusions}
In this work we have shown that interference of the gluon fusion Higgs production channel $gg \to \hgll$ with the background $2 \to 3$ process $\glgll$ gives rise to an observable that is unambiguously sensitive to the parity structure of the $\hzgver$ vertex. This observable manifests as an amplitude modulation and phase shift of the oscillatory angular probability distribution~\eqref{eqn:MRP} with respect to the azimuthal angle, $\phi_Z$. Equivalently, the parity violation manifests respectively as a rescaling of the cross-section and a parity asymmetry on $\phi_Z$ quadrants. However, only the phase shift and its associated quadrant asymmetry are theoretically clean observables. 

Numerical simulations with MCFM, privately modified to include Higgs-background interference in the $\glgll$ channel, estimate that for the SM Higgs, the quadrant asymmetry (cross-section rescaling) enters at the $10^{-3}$ ($10^{-2}$) level for $\xi \sim 1$. Unfortunately, the large $q\bar{q}$ incoherent background, combined with the very narrow Higgs width, renders this background interference effect too small to be seen at the LHC, even for a future high luminosity of $3$~ab$^{-1}$.

The analysis in this paper, however, generically holds for any scalar that may be produced by gluon fusion and has a decay channel to $Z\gamma$. If there exists a new scalar with either a larger gluon fusion production cross--section times $Z\gamma$ branching ratio or a larger total width, then the parity structure of this coupling may be probed or constrained by searches for a $\phi_Z$ phase shift or quadrant asymmetry.

\section*{Acknowledgements}
We thank Jessie Shelton and Jon Walsh for useful discussions. We particularly thank Jon Walsh for his assistance with numerical simulations. The work of MF and YG is supported in part by the U.S.\ National Science Foundation through grant PHY-0757868. The work of YG is also supported by the United States-Israel Binational Science Foundation (BSF) under grant no.~2010221. The work of DR is supported by the NSF under grant No. PHY-1002399.

%%%%%%%%%%%%%%%%%%%%%%%%%%%%%

\appendix
\section{Phase Space Construction}
\label{app:KV}
Here we provide explicit expressions for the phase space coordinates $\{s,\thg;\thz,\phiz\}$ in terms of kinematic observables. First, with respect to the digluon invariant mass $(k_1 + k_2)^2 = s$ and $\thg$, the other two $2 \to 2$ Mandelstam variables
\begin{equation}
	t = (k_1-k)^2 = (\mz^2 - s)\sin^2(\thg/2)~, \quad \mbox{and} \quad  u = (k_2 - k)^2 = (\mz^2 - s)\cos^2(\thg/2)~.
\end{equation}	
Hence $t-u = (s-\mz^2) \cos(\thg)$, i.e.
\begin{equation}
	\thg = \cos^{-1}\bigg[\frac{t-u}{s-\mz^2}\bigg]~,
\end{equation}
on the branch $\thg \in [0,\pi]$. Note that since $s + t + u = \mz^2$ in the on-shell $Z$ limit, the $2 \to 2$ amplitudes may always be expressed as functions of $s$ and $t-u = (s-\mz^2) \cos(\thg)$ alone. That is, we see explicitly here that the $2 \to 2$ phase space is fully specified by $s$ and $\thg$.

The polar angle $\thz$ is similarly extracted by noting that $2(p_+ - p_-)\cdot k = (s - \mz^2)\cos\thg$, i.e.
\begin{equation}
	\thz = \cos^{-1}\bigg[\frac{2(p_+ - p_-)\cdot k}{s-\mz^2}\bigg]~.
\end{equation}
Finally, the Levi-Civita contraction
\begin{equation}
	\label{eqn:ALCC}
	\varepsilon^{\mu\nu\rho\sigma}{k_1}_\mu {k_2}_\nu k_\rho {p_+}_\sigma = \frac{1}{8}\mz\sqrt{s}(s-\mz^2)\sin(\thg)\sin(\thz)\sin(\phiz)~,
\end{equation}
and
\begin{equation}
	\label{eqn:AGLC}
 	2(k_1 - k_2)\cdot(p_+ - p_-)  = (s + \mz^2)\cos(\thg) \cos(\thz)  - 2 \sqrt{s}\mz \sin(\thg) \sin(\thz) \cos(\phiz)~.
\end{equation}
Assuming only longitudinal gluon boosts, so that the gluons are oriented along the beam line, then the $k_{1,2}^\mu$ lab frame components may be extracted from $s$ and the total energy of the outgoing states. Then, the relations~\eqref{eqn:ALCC} and \eqref{eqn:AGLC} permit extraction of $\sin\phiz$ and $\cos\phiz$, and hence $\phiz \in [0,2\pi]$ without any ambiguities.

\section{Conventions}
\label{app:SPC}
Our choices for polarizations and reference momenta are
\begin{gather}
	\epsilon^\pm_\mu(k_i) = \pm\frac{\sigbr{\mp}{k}{\sigma_\mu}{k_i}}{\sqrt{2}\langle k^\mp | k_i^\pm\rangle}~, \qquad \epsilon^\pm_\mu(k) = \pm\frac{\sigbr{\mp}{\tilde{p}}{\sigma_\mu}{k}}{\sqrt{2}\langle \tilde{p}^\mp | k^\pm\rangle}\notag\\
	\epsilon^{\pm}_\mu(p) = \pm\frac{\sigbr{\mp}{k}{\sigma_\mu}{\tilde{p}}}{\sqrt{2}\langle k^\mp | \tilde{p}^\pm\rangle}~, ~~\epsilon^{0}_\mu(p) = \frac{\tilde{p}_\mu}{\mz} - \frac{\mz k_\mu}{2 k\cdot \tilde{p}}~, \label{eqn:PCV}
\end{gather}
where $p$ is the $Z$ momentum, and $\tilde{p}^\mu = p^\mu - \mz^2 k^\mu/(2k\cdot p)$ is a null associated momentum, such that $\tilde{p}$ and $k$ form a light-cone decomposition of $p$. We assume the leptons are massless, and make the spinor phase choices
\begin{equation}
	\label{eqn:SPC}
	\lambda^a_q = \begin{cases} \Big( \sqrt{q^0 + q^3}, (q^1 - i q^2)/\sqrt{q^0 + q^3}\Big)~, & q = p_{\pm}, \tilde{p}, k_1~, \\ \Big((q^1 + i q^2)/\sqrt{q^0 - q^3},\sqrt{q^0 - q^3}\Big)~, & q = k, k_2~.\end{cases}
\end{equation}

\section{Parity of \texorpdfstring{$2 \to 2$}{22} amplitudes}
\label{app:PHA}
To deduce the parity of $[\mathcal{M}_{2 \to 2, \rm{bg}}]^{\lambda_1\lambda_2}_{\lambda\kappa}$, we may consider the heavy quark limit. In this case, the external spin states determine the total angular momentum, such that the $\thg$ dependence of each helicity amplitude is encoded by Wigner d-matrix functions, viz.
\begin{equation}
	\mathcal{M}_{2 \to 2,\rm{bg}}(s,\thg) = f(s) \langle j;m'| e^{-i\thg J_y} |j;m\rangle \equiv f(s) d^j_{m,m'}(\thg)~,
\end{equation}
for some $f$, where $m$ and $m'$ are spin projections. It follows that
\begin{equation}
	[\mathcal{M}_{2 \to 2,\rm{bg}}]^{\lambda_1 \lambda_2}_{\lambda \kappa}(\thg)  \sim d^j_{\lambda_1 - \lambda_2, \lambda-\kappa}(\thg)~,
\end{equation}
where $2 \le j \le \mbox{max}\{\lambda_1-\lambda_2,\lambda-\kappa\}$. The parity structure of the Wigner d-matrix functions immediately implies that the parity of these amplitudes is $-(-1)^\kappa$, since $\lambda_i$, $\lambda = \pm 1$ only. That is,
\begin{equation}
	[\rm{P}\mathcal{M}_{2 \to 2,\rm{bg}}]^{\lambda_1 \lambda_2}_{\lambda \kappa}(s,\thg) = -(-1)^\kappa [\mathcal{M}_{2 \to 2,\rm{bg}}]^{\lambda_1 \lambda_2}_{\lambda \kappa}(s,\thg)~.
\end{equation}
This matches the Higgs channel result~\eqref{eqn:HCPR}, which are similarly generated by local effective operators, and also the massless quark results in Ref.~\cite{Ametller:1985ro}.

%\bibliography{../hZgamma}


\end{document}